\documentstyle[12pt]{article}
\def\lsim{\mathrel{\lower2.5pt\vbox{\lineskip=0pt\baselineskip=0pt
           \hbox{$<$}\hbox{$\sim$}}}}
\def\gsim{\mathrel{\lower2.5pt\vbox{\lineskip=0pt\baselineskip=0pt
           \hbox{$>$}\hbox{$\sim$}}}}
\newcommand{\drawsquare}[2]{\hbox{%
\rule{#2pt}{#1pt}\hskip-#2pt%  left vertical
\rule{#1pt}{#2pt}\hskip-#1pt%  lower horizontal
\rule[#1pt]{#1pt}{#2pt}}\rule[#1pt]{#2pt}{#2pt}\hskip-#2pt%  upper horizontal
\rule{#2pt}{#1pt}}% right vertical
\newcommand{\Yfund}{\raisebox{-.5pt}{\drawsquare{6.5}{0.4}}}%  fund
\newcommand{\Yasymm}{\raisebox{-3.5pt}{\drawsquare{6.5}{0.4}}\hskip-6.9pt%
        \raisebox{3pt}{\drawsquare{6.5}{0.4}}}%  antisymmetric second rank
\begin{document}
\setlength{\baselineskip}{8mm}
\begin{titlepage}
\begin{flushright}
\begin{tabular}{c c}
%& {\normalsize  hep-ph/9609395} \\
%& {\normalsize  DPNU-96-49} \\
& {\normalsize July 1998}
\end{tabular}
\end{flushright}
\vspace{5mm}
\begin{center}
{\large \bf Composite model with neutrino large mixing}\\
\vspace{15mm} 
Naoyuki Haba\footnote{E-mail:\ haba@eken.phys.nagoya-u.ac.jp : 
Address after Sept. 1, (98): 
Department of Physics, Ohio State University, 
174 W. 18th Ave., Columbus, Ohio, 43210}\\

\vspace{5mm}

{\it Faculty of Engineering, Mie University, Mie, Japan 514-0008 \\
                and \\
Department of Physics, Nagoya University, Nagoya, Japan 464-8602 \\
}
\end{center}

\vspace{10mm}

%%%%%%%%%%%%%%%%%%%%%%%%%%%%%%%%%%%%%%%%%%%%%%%%%%%%%%%%%%%%%%%%%%%
%%%%%%%%%%%%%%%%%%%%%%%%%%%%%%%%%%%%%%%%%%%%%%%%%%%%%%%%%%%%%%%%%%%
%%%%%%%%%%%%%%%%%%%%% ABSTRACT %%%%%%%%%%%%%%%%%%%%%%%%%%%%%%%%%%%%
%%%%%%%%%%%%%%%%%%%%%%%%%%%%%%%%%%%%%%%%%%%%%%%%%%%%%%%%%%%%%%%%%%%
%%%%%%%%%%%%%%%%%%%%%%%%%%%%%%%%%%%%%%%%%%%%%%%%%%%%%%%%%%%%%%%%%%%

\begin{abstract}

We suggest a simple composite model that induces 
the large flavor mixing of neutrino in the supersymmetric theory. 
This model has only one hyper-color in addition
to the standard gauge group, 
which makes composite states of preons. 
In this model, {\bf 10} and {\bf 1} representations 
in $SU(5)$ grand unified theory are composite states 
and produce the mass hierarchy. 
This explains why the large mixing is realized 
in the lepton sector, while the small mixing is realized in the 
quark sector.
This model can naturally solve 
the atmospheric neutrino problem. 
We can also solve the solar neutrino problem 
by improving the model.

\end{abstract}
\end{titlepage}

%%%%%%%%%%%%%%%%%%%%%%%%%%%%%%%%%%%%%%%%%%%%%%%%%%%%%%%%%%%%%%%%%%%
%%%%%%%%%%%%%%%%%%%%%%%%%%%%%%%%%%%%%%%%%%%%%%%%%%%%%%%%%%%%%%%%%%%
%%%%%%%%%%%%%%%%%%%%% INTRODUCTION %%%%%%%%%%%%%%%%%%%%%%%%%%%%%%%%
%%%%%%%%%%%%%%%%%%%%%%%%%%%%%%%%%%%%%%%%%%%%%%%%%%%%%%%%%%%%%%%%%%%
%%%%%%%%%%%%%%%%%%%%%%%%%%%%%%%%%%%%%%%%%%%%%%%%%%%%%%%%%%%%%%%%%%%
%
\section{Introduction}

Recent experiments of Superkamiokande suggest 
the large mixing between $\nu_{\mu}$ and other
neutrinos\cite{1}\cite{1.5}.
As for the solar neutrino problem\cite{2}, 
there is the large angle solution 
of the matter induced resonant Mikheyev-Smirnov-Wolfenstein 
(MSW) oscillation\cite{MSW} between $\nu_e$ 
and other neutrinos as well as 
the small angle MSW solution\cite{Hata}. 
The vacuum oscillation solution\cite{vac} 
also suggests the large mixing of 
neutrinos\footnote{Recent Superkamiokande data of 
the electron energy spectrum 
suggest the vacuum oscillation solution with 
maximal mixing 
is favored\cite{6}\cite{7}. 
}.

\par
These experimental results seem to suggest 
the possibility that 
the large flavor mixing
is realized in the lepton sector\footnote{
Although the LSND results suggest the small mixing 
between $\overline{\nu}_{\mu}$ and 
$\overline{\nu}_{e}$\cite{LSND},
confirmation of the LSND results still awaits future
experiments. 
Recent measurements in the KARMEN
detector exclude part of the LSND allowed region\cite{KARMEN}.
}. 
If it is true, 
one has to explain 
why the large mixing is realized 
in the lepton sector, while the small
mixing is realized in the quark sector. 
The possibility of neutrino large mixing 
has been studied from the view point of the mass matrix
texture\cite{Fukugita}\cite{10.5}, 
the sea-saw enhancement mechanism\cite{texture}, 
the analysis of the renormalization group equation\cite{RGE},
the grand unified theories\cite{Barr}\cite{Barr2},
the pseudo-Dirac neutrino mass\cite{pD}, 
the singular see-saw mechanism\cite{sing},
the radiative induced neutrino mass\cite{Ma},
and so on.

\par
In this paper 
we suggest a simple model that 
naturally induces the large
mixing of neutrinos in the supersymmetric gauge theory. 
This model has only one hyper-color in addition
to the standard gauge group, 
which makes composite states of preons. 
If {\bf 10} and {\bf 1} representations 
in $SU(5)$ grand unified theory (GUT) 
are composite states, 
they produce mass hierarchy 
which induces the large mixing 
in the lepton sector and the small mixing in the 
quark sector. 
This model can naturally solve 
the atmospheric neutrino problem. 
We can also solve the solar neutrino problem 
by improving the model.

\par
In section 2, we suggest a composite model. 
In section 3, we try to improve the model in order 
to solve the solar neutrino problem. 
Section 4 gives summary and discussions.

%
%%%%%%%%%%%%%%%%%%%%%%%%%%%%%%%%%%%%%%%%%%%%%%%%%%%%%%%%%%%%%%%%%%%%%%%%%
%%%%%%%%%%%%%%%%%%%%%%%%%%%%%%%%%%%%%%%%%%%%%%%%%%%%%%%%%%%%%%%%%%%%%%%%%
%%%%%%%%%%%%%%%%%%%%%%%%%%%%%  Section 2  %%%%%%%%%%%%%%%%%%%%%%%%%%%%%%%
%%%%%%%%%%%%%%%%%%%%%%%%%%%%%%%%%%%%%%%%%%%%%%%%%%%%%%%%%%%%%%%%%%%%%%%%%
%%%%%%%%%%%%%%%%%%%%%%%%%%%%%%%%%%%%%%%%%%%%%%%%%%%%%%%%%%%%%%%%%%%%%%%%%
%
\section{A composite model}

In the supersymmetric gauge theory,
various composite models have been built 
in the s-confinement theory\cite{Strassler}\cite{s-confine}.  
One example of 
the s-confinement theory is the 
$Sp(2 N)$ gauge theory with one antisymmetric 
tensor $A$ and six fundamentals $Q$s\cite{st1}.
Kaplan, Lepeintre and Schmaltz have built 
composite models
by using this theory\cite{st2}.
In this paper, we try to build the composite model 
of $SU(5)$ GUT by using this theory.

\par
We consider $SU(5)$ GUT with 
three right-handed neutrinos $\overline{N_R^c}$s. 
Quarks and leptons are represented by 
${\bf 10}_i$, ${\overline{\bf 5}_i}$, and ${{\bf 1}_i}$
representations of $SU(5)$ as 
\begin{eqnarray}
\label{miracle}
& & {\bf 10}_i = ( Q_L, 
       \overline{U_R^c}, \overline{E_R^c} )_i \;, \nonumber \\
& & {\overline{\bf 5}_i} = ( \overline{D_R^c}, L_L  )_i \;, \\
& & {\bf 1}_i = ( \overline{N_R^c} )_i \;, \nonumber
\end{eqnarray}
where the index $i$ ($i = 1,2,3$) stands for 
the generation number. 
$Q_L$, $L_L$, $\overline{U_R^c}$, $\overline{D_R^c}$, and 
$\overline{E_R^c}$ express quark doublet, lepton doublet, 
right-handed up-sector, right-handed 
down-sector and right-handed 
charged lepton fields,
respectively.

\par
The $Sp(2N)$ theory with one $A$ and six $Q$s 
has the composite states 
\begin{eqnarray}
  {\rm Tr} A^m, & & m = 2, 3, ..., N, \nonumber \\
  Q A^n Q,      & & n = 0, 1, ..., N-1   \; .
\end{eqnarray}
We consider the case of $N = 3$.
We assume that the field $Q$ transforms 
under the $SU(5)$ gauge symmetry as well as 
$Sp(6)$ gauge symmetry, and 
the gauge coupling of $SU(5)$ 
is much weaker than that of $Sp(6)$.
In our model we introduce preon fields 
as follows. 
\begin{equation}
\label{preons}
 \begin{array}{c|cccc}
  {\rm preon}   &   Sp(6)      &    SU(5)   & (U(1)_w) & Z_2   \\ 
\hline
  & &  &  & \\
  A             &   \Yasymm    &      1    &   &    \\
  Q             &   \Yfund     &  \Yfund   &   &    \\ 
  Q'            &   \Yfund     &      1    &   & -  \\
 \overline{Q_i} &     1        & \overline{\Yfund} &  &   \\  
 \overline{\chi_j} &  1        & \overline{\Yfund} &  & - \\
  H             &     1        &  \Yfund           & (+1) & \\ 
 \overline{H}   &     1        & \overline{\Yfund} &  &   \\  
  N_3           &     1        &      1    &   &    \\        
  S             &     1        &      1    &  (-1)    &  \\        
 \end{array}
\end{equation}  
$\overline{Q_i}$ is 
the matter field of ${\overline{\bf 5}_i}$ in 
Eq.(\ref{miracle}). 
$\overline{\chi_j}$ $(j=1 \sim 3)$ is 
introduced in order to cancel the gauge anomaly, 
where the index $j$ has no relation to 
the generation number. 
$H$ and $\overline{H}$ are Higgs fields 
which are singlet under $Sp(6)$ gauge symmetry. 
We introduce $Z_2$ discrete symmetry 
in order to distinguish Higgs fields 
$H, \overline{H}$ and 
matter field $\overline{Q_i}$ from extra fields. 
$N_3$ is the right-handed neutrino of the third generation. 
$S$ is singlet under both $Sp(6)$ and $SU(5)$ 
gauge symmetries. 
In addition to $Sp(6)$ and $SU(5)$ gauge symmetries, 
we introduce anomalous $U(1)_w$ gauge symmetry\footnote{
We expect the $U(1)_w$ anomaly is canceled 
by the Green-Schwarz mechanism\cite{GS}.
}, 
which induces a Fayet-Iliopoulos term 
$\xi^2 \sim (g_S^2 / 192 \pi^2) \:{\rm Tr}\: q_w \: M_p^2$ from 
the string loop corrections\cite{FI}, 
where $g_S$ and $q_w$ are the string 
coupling and the charge of the 
anomalous $U(1)_w$ gauge symmetry, 
respectively.  
$S$ and $H$ have charges of $U(1)_w$ 
as Eq.(\ref{preons}). 
We assume that many extra fields $X$s 
which have plus $U(1)_w$ charges 
induce a Fayet-Iliopoulos term $\xi$ of 
order the Planck scale. 
Then we can expect 
that $S$ obtains the vacuum 
expectation value (VEV) of order  
the Planck scale $M_p \simeq O(10^{18})$ GeV. 
Extra fields $X$s do not contribute to 
the low energy phenomenology and 
mass matrices of quark and lepton. 
We will see the reason why 
the field $S$ and $U(1)_w$ gauge symmetry 
are introduced later.

\par
We consider the situation that $Sp(6)$ 
dynamical scale 
$\Lambda$ satisfies $M_{GUT} < \Lambda < M_p$, 
where $M_{GUT}$ is the $SU(5)$ 
GUT scale of $O(10^{16})$ GeV. 
This condition is required by 
the proton stability. 
If $\Lambda < M_{GUT}$, 
$D$-term interactions in K\"ahler potential 
which are suppressed 
by $(1/ \Lambda)$ might induce too rapid proton 
decay\footnote{
I would like to thank Prof. T. Yanagida for teaching me this process.
}.

\par
Bellow the scale of $\Lambda$,
this theory is described 
by the $Sp(6)$ singlet states. 
We regard these $Sp(6)$ singlet states 
as quark, lepton, 
Higgs, and extra fields as follows. 
\begin{equation}
\label{comp}
 \begin{array}{l|ccc}
                    &    SU(5)    &  (U(1)_w)  &   Z_2       \\ 
\hline
  & & &  \\
  N_1 = A^3         &   {\bf 1}   & &         \\
  N_2 = A^2         &   {\bf 1}   & &         \\
  N_3               &   {\bf 1}   & &         \\
{\bf 10}_1 = QA^2Q  &   \Yasymm   & &         \\
{\bf 10}_2 = QAQ    &   \Yasymm   & &         \\
{\bf 10}_3 = Q^2    &   \Yasymm   & &         \\
  \chi_1   = Q Q'   &   \Yfund    & & -        \\
  \chi_2   = QAQ'   &   \Yfund    & & -         \\ 
  \chi_3   = QA^2Q' &   \Yfund    & & -         \\
{\overline{\bf 5}_i}= \overline{Q_i} & \overline{\Yfund} & &  \\
 \overline{\chi_j}  & \overline{\Yfund}    & & -    \\ 
  H                 &    \Yfund            & (+1) &   \\ 
 \overline{H}       &    \overline{\Yfund} & &      \\
  S                 &   {\bf 1}            & (-1) &   \\  
 \end{array}
\end{equation}  

\par
The mass term of Higgs particles 
$H$ and $\overline{H}$ 
is induced from the operator 
$[\langle S \rangle H \overline{H}] \sim M_p H \overline{H}$. 
We assume that triplet-doublet splitting 
is realized in another mechanism, 
and Higgs doublets obtain 
suitable weak scale VEVs 
as $\langle H \rangle = v$ and 
$\langle \overline{H} \rangle = \overline{v}$, 
where $v^2 + \overline{v}^2 = (174 \ {\rm GeV})^2$, 
according to 
the supersymmetry (SUSY) breaking effects\footnote{
In this paper 
we assume that the SUSY is broken at the low energy, 
whose effects are negligible at the scale of $M_{GUT}$.
}.

\par
We assume $\chi_j$ and $\overline{\chi_j}$ 
do not take VEVs. 
$Z_2$ symmetry distinguishes $\chi_j$ from 
$H$, and $\overline{\chi_j}$ from 
$\overline{H}, \overline{Q_i}$. 
At the scale of $\Lambda$, 
the mass matrix of $\chi_j$ and $\overline{\chi_j}$ 
is given by 
$$
\left( 
    \begin{array}{ccc}
      \overline{\chi_3} & \overline{\chi_2} & \overline{\chi_1} \\
    \end{array}
  \right) \;
\left(
\begin{array}{ccc}
  \epsilon    &  \epsilon^2    &  \epsilon^3      \\
  \epsilon    &  \epsilon^2    &  \epsilon^3      \\
  \epsilon    &  \epsilon^2    &  \epsilon^3      \\
\end{array}
\right) \; M_p \;
\left( 
    \begin{array}{c}
      \chi_1        \\
      \chi_2        \\
      \chi_3        \\
   \end{array}
  \right) \; ,
$$
where $\epsilon \equiv \Lambda/M_p$. 
{}For example, 
the mass of $\chi_1$ and $\overline{\chi_1}$ 
is induced as 
$[(QQ') \overline{\chi_1}] 
\sim \Lambda \chi_1 \overline{\chi_1}$.
The dimension-less parameter 
$\epsilon$ is very important which will express the 
mass hierarchy of quark and lepton later. 
The above mass matrix has 
$O(1)$ coefficients, 
and we do not consider the case 
of zero determinant. 
Then three mass eigen values of 
$\chi_j$ and $\overline{\chi_j}$ 
are of $O(\epsilon M_p)$, $O(\epsilon^2 M_p)$, 
and $O(\epsilon^3 M_p)$. 
They are heavy enough not to affect the low energy 
phenomenology.

\par
The $Sp(6)$ strong dynamics induces 
the non-perturbative superpotential 
$ W_{dyn}$\cite{st1}, which is 
written as 
\begin{eqnarray}
 \label{Wdyn2}
  W_{dyn} &\simeq& 
    {g^3 \over  \Lambda^2} N_2^2 {\bf 10}_2^2 \chi_1 +
    {g^2 \over  \Lambda  } N_1 [ {\bf 10}_1 {\bf 10}_2 \chi_1 +
                              \chi_2 {\bf 10}_2  {\bf 10}_2  ]  \nonumber \\
& &+{g^2 \over  \Lambda  } N_2 [ {\bf 10}_3 {\bf 10}_2 \chi_1 +
                              \chi_3 {\bf 10}_2  {\bf 10}_2  ] \nonumber \\  
& &+{g}[ {\bf 10}_1 {\bf 10}_1 \chi_3 + 
                             {\bf 10}_1 {\bf 10}_3 \chi_2 ]
   +{g}[ {\bf 10}_3 {\bf 10}_3 \chi_1    + 
                             {\bf 10}_3 {\bf 10}_2 \chi_3 ] 
\end{eqnarray}
according to the field assignment of Eq.(\ref{comp}). 
Here the factor $g \simeq 4 \pi$ 
follows the power counting arguments in Ref.\cite{NDA}.
Since $\chi_j$s have masses around GUT scale and 
do not take VEVs, 
the dynamically generated interactions of 
Eq.(\ref{Wdyn2}) have nothing to do 
with the low energy phenomenology.

\par
Under the above assumptions, 
masses of quark and lepton are produced from 
irrelevant operators suppressed by the Planck scale. 
Yukawa interactions which include 
composite states ({\bf 10} and {\bf 1}) 
are suppressed by the dimension-less parameter $\epsilon$. 
The mass hierarchy is generated since quark and lepton 
are composite states. 
This model is one of the models 
of Froggatt-Nielsen mechanism\cite{FN}.
In order to obtain the suitable mass hierarchy,
we set $\epsilon \sim 1/10$, 
which suggests $g \sim 1/ \epsilon$. 
We denote $3 \times 3$ flavor space 
mass matrices as $m_{ij}$, which represents 
$L_i m_{ij} R_j$, 
where $L_i$ and $R_j$ are left- and 
right-handed fermions, respectively. 
The mass matrix of up-sector $m^{u}_{ij}$, 
down-sector $m^{d}_{ij}$, 
charged lepton $m^{e}_{ij}$, and 
Dirac neutrino $m^{\nu}_{ij}$ are given by 
\begin{equation}
\label{up2}
m^{u}_{ij} \simeq 
 \left(
\begin{array}{ccc}
\epsilon^6   &  \epsilon^5    & \epsilon^4      \\
\epsilon^5   &  \epsilon^4    & \epsilon^3      \\
\epsilon^4   &  \epsilon^3    & \epsilon^2      \\
\end{array}
 \right) \: g^2 \: v  
\sim 
 \left(
\begin{array}{ccc}
\epsilon^4   &  \epsilon^3    & \epsilon^2      \\
\epsilon^3   &  \epsilon^2    & \epsilon        \\
\epsilon^2   &  \epsilon      & 1               \\
\end{array}
 \right) \: v ,
\end{equation}
\begin{equation}
\label{down2}
m^{d}_{ij} \simeq 
 \left(
\begin{array}{ccc}
 \epsilon^3   &  \epsilon^3   &  \epsilon^3    \\
 \epsilon^2   &  \epsilon^2   &  \epsilon^2    \\
 \epsilon     &  \epsilon     &  \epsilon      \\
\end{array}
 \right) \: g \: \overline{v}  
\sim 
 \left(
\begin{array}{ccc}
 \epsilon^2   &  \epsilon^2   &  \epsilon^2    \\
 \epsilon     &  \epsilon     &  \epsilon      \\
 1            &  1            &  1             \\
\end{array}
 \right) \: \overline{v},
\end{equation}
\begin{equation}
\label{lepton2}
m^{e}_{ij} \simeq 
 \left(
\begin{array}{ccc}
 \epsilon^3   &  \epsilon^2     &  \epsilon     \\
 \epsilon^3   &  \epsilon^2     &  \epsilon     \\
 \epsilon^3   &  \epsilon^2     &  \epsilon     \\
\end{array}
 \right) \: g \: \overline{v}  
\sim 
 \left(
\begin{array}{ccc}
 \epsilon^2   &  \epsilon     &  1     \\
 \epsilon^2   &  \epsilon     &  1     \\
 \epsilon^2   &  \epsilon     &  1     \\
\end{array}
 \right)  \: \overline{v},
\end{equation}
\begin{equation}
\label{neuD2}
m^{\nu}_{ij} \simeq 
 \left(
\begin{array}{ccc}
  \epsilon^2   & \epsilon   &  1   \\
  \epsilon^2   & \epsilon   &  1   \\
  \epsilon^2   & \epsilon   &  1   \\
\end{array}
 \right) \: g^2 \: v  ,
\end{equation}
respectively. 
{}For example, 
$m^{u}_{11}$ is induced from the 
operator $[{g^2 \over M_p^7}(QA^2Q)(QA^2Q)HS]$. 
The Dirac masses of 
$m^{u}_{ij}$ and $m^{\nu}_{ij}$ have factor 
$g^2$ while $m^{d}_{ij}$ and $m^{e}_{ij}$ 
have factor $g$. 
It is because $m^{u}_{ij}$ and $m^{\nu}_{ij}$ 
have additional factor 
$ g \langle S \rangle/M_p \sim g$. 
That is why 
we introduce the field $S$ and 
extra $U(1)_w$ gauge symmetry\footnote{
Although $S$ takes the VEV of $O(M_p)$, 
higher order operators in the superpotential 
$[(S H)^n (S^k X^l)^m]$ 
are negligible. 
It is because $\langle H \rangle \simeq 10^2$ GeV and 
$\langle X \rangle = 0$. 
These operators have nothing to do with the 
low energy phenomenology and 
mass matrices of quark and lepton. 
}. 
The Majorana mass of right-handed neutrinos $M_{\nu}$
is given by
\begin{equation}
\label{majo22}
M_{\nu} \simeq 
\left(
\begin{array}{ccc}
  \epsilon^4     &  \epsilon^3   &  \epsilon^2     \\
  \epsilon^3     &  \epsilon^2   &  \epsilon       \\
  \epsilon^2     &  \epsilon     &  1              \\
\end{array}
\right) \: M_p .
\end{equation}
Through the see-saw mechanism
the mass matrix of three light 
neutrinos $m_{\nu_l}$ becomes 
\begin{eqnarray}
\label{62}
& &  m_{\nu_l} 
  = - m_{\nu} M_{\nu}^{-1} m_{\nu}{}^T , \nonumber \\
& & \simeq  
 - \left( 
    \begin{array}{c}
     1  \\
     1  \\
     1  \\
   \end{array}
  \right) 
\left(
\begin{array}{ccc}
  \epsilon^2   &  \epsilon  &  1    \\
\end{array}
\right) 
\left(
\begin{array}{ccc}
1/ \epsilon^4          &  1/ \epsilon^3   &  1/ \epsilon^2   \\
1/ \epsilon^3          &  1/ \epsilon^2   &  1/ \epsilon     \\
1/ \epsilon^2          &  1/ \epsilon     &       1          \\
\end{array}
\right) 
\left( 
    \begin{array}{c}
     \epsilon^2    \\
     \epsilon      \\
     1             \\
   \end{array}
  \right)
\left(
\begin{array}{ccc}
  1 & 1 & 1 \\
\end{array}
\right) {g^4 v^2 \over M_p},  \nonumber \\ 
& & \simeq  
 - \left( 
 \begin{array}{ccc}
  1 & 1 & 1 \\
  1 & 1 & 1 \\
  1 & 1 & 1 \\
 \end{array}
 \right) {g^4 v^2 \over M_p} .  
\end{eqnarray}
Above mass matrices 
shows only order, and all elements 
have $O(1)$ coefficients. 
Here we consider the situation that 
all matrices have non-zero determinant\footnote{
If the determinant of 
Majorana mass of right-handed neutrino of Eq.(\ref{majo22}) 
is zero, the inverse matrix $M_{\nu}^{-1}$
does not exist. 
However, it is interesting to consider such the 
case, because 
the singular see-saw mechanism can work\cite{sing}.
}.

\par
This model induces the following mass hierarchy.
\begin{eqnarray}
\label{ratio2}
& & m_d : m_s : m_b \sim  m_e : m_{\mu} : m_{\tau} \sim 
    \epsilon^2 : \epsilon : 1   \nonumber \\ 
& & m_u : m_c : m_t \sim 
    \epsilon^4 : \epsilon^2 : 1 
\end{eqnarray}
It suggests almost realistic mass hierarchy\footnote{
The mass hierarchy of Eq.(\ref{ratio2}) is not correct for 
the first and the second generations of 
the down-quark and charged lepton sectors, since 
experiments suggest 
$m_d/m_s \sim m_e/m_{\mu}\sim O(10^{-2})$. 
In the next section, 
we will try to modify 
Eq.(\ref{ratio2}) 
by improving the model. 
} 
with $\epsilon \sim 1/10$.  
This model suggests the large 
$\tan \beta \; (\equiv v/ \overline{v})$.

\par
Let us show that this model 
naturally induces the large mixing 
in the lepton sector 
and the small mixing 
in the quark sector. 
{}For the quark sector, 
mass matrices $m^u_{ij}$ and $m^d_{ij}$ 
in Eqs.(\ref{up2}) and (\ref{down2}) derives 
\begin{equation}
\label{KMq}
V_{KM}^{\rm quark}{}_{i(i+n)}\equiv U^u_L{}^{\dagger}U^d_L
\simeq \epsilon^n , 
\;\;\;\;\; \; \; \; \; n=0, 1, 2 \; ,
\end{equation}
where $U^u_L$ and $U^d_L$ are 
unitary matrices which diagonalize 
$m^u_{ij}$ and $m^d_{ij}$ from 
the left-hand side, respectively. 
Then we can predict 
\begin{equation}
\label{14}
V_{KM}^{\rm quark}{}_{ij}\simeq m^d_i/m^d_j  
\end{equation}
{}from Eqs. (\ref{ratio2}) and (\ref{KMq}). 
If we input experimental values of masses 
in Eq.(\ref{14}), we obtain
\begin{eqnarray}
\label{VKMq}
& & V_{us} \simeq {m_d^{(exp)} \over m_s^{(exp)}} 
           \sim 0.03 \sim 0.07 \; , \;\;\;\;\;\;\;
    V_{cb} \simeq {m_s^{(exp)} \over m_b^{(exp)}} 
           \simeq 0.02 \sim 0.04 \; , \\
& & {V_{ub} \over V_{cb}} \simeq 
     {m_d^{(exp)}/m_b^{(exp)} \over m_s^{(exp)}/m_b^{(exp)}} 
\sim 0.03 \sim  0.07 \; , \nonumber
\end{eqnarray}
where we use $m^{(exp)}$ as the mass at 
1 GeV\cite{25.5}, for one example\footnote{
The explicit values of $V_{KM}^{\rm quark}$ 
in Eq.(\ref{VKMq}) have no meaning. 
Here we would like to discuss 
order of the values of 
$V_{KM}^{\rm quark}$ elements. 
}. 
This naive estimation derives too small $V_{us}$ 
compared to the experimental value 
$V_{us}^{(exp)} \simeq 0.22$. 
It is because we used wrong 
mass hierarchy $m_d/m_s \sim 10^{-1}$ of Eq.(\ref{ratio2}) 
when we estimate Eq.(\ref{VKMq}). 
Experiments suggest 
$m_d^{(exp)}/m_s^{(exp)} \sim 10^{-2}$ as Eq.(\ref{VKMq}). 
On the other hand, 
$V_{cb}$ and $V_{ub}/V_{cb}$ are roughly consistent 
with experimental values of 
$V_{cb}^{(exp)} \simeq 0.036 \sim 0.046$ 
and 
$V_{ub}^{(exp)}/V_{cb}^{(exp)} \simeq 0.04 \sim 0.14$.

\par
How about the lepton flavor mixing ? 
Equations (\ref{lepton2}) and (\ref{62}) 
suggest that unitary matrices $U_L^e$ and $U^{\nu}$, 
which diagonalize $m^e_{ij}$ 
and $m_{\nu}$ from the left-hand side\footnote{
Since $m_{\nu}$ is Hermite, 
it is diagonalized by $U^{\nu} m_{\nu} U^{\nu}{}^T$. 
}, respectively, 
have both the same form as 
\begin{equation}
\label{UU}
 U_L^e \sim U^{\nu} \simeq  \left( 
 \begin{array}{ccc}
  O(1) & O(1) & O(1) \\
  O(1) & O(1) & O(1) \\
  O(1) & O(1) & O(1) \\
 \end{array}
\right) .
\end{equation}
Thus, the neutrino mixing matrix 
$V^{\rm lepton}_{}\equiv U_L^e{}^{\dagger}U^{\nu}$ 
suggests the large mixing of $O(1)$, 
as long as the accidental cancellation 
occurs between $U_L^e$ and $U_{\nu}$. 
Equation (\ref{62}) suggests 
mass squared differences are of order 
$\delta m^2 \simeq \mu^2 \simeq 10^{-2}$ 
eV$^2$ $(\mu \equiv g^4 v^2/M_p)$, 
which can be the solution of atmospheric neutrino 
problem with large mixing between 
$\nu_{\mu}$ and $\nu_{\tau}$ of 
$\delta m_{23}^2 \simeq 4 \sim 6 \times 10^{-3}$ 
eV$^2$\cite{1}\footnote{
To be accurate, 
we need to know the 
coefficients in Eqs.(\ref{lepton2}) and (\ref{62}) 
to check whether this $O(1)$ mixing is maximal or not. 
However, this model can not predict 
the coefficients. 
}.
{}From Eq.(\ref{62}) we can see that 
the large mixing has nothing to do 
with the explicit form of Majorana mass at all. 
The Dirac mass structures of Eqs.(\ref{lepton2}) 
and (\ref{neuD2}) are crucial
{}for this large mixing, 
because 
the mass hierarchy of charged lepton and Dirac neutrino 
are produced only by right-handed fields. 
The flavor mixing is determined by the unitary matrices 
which diagonalize mass matrices from the left-hand side.

\par
In this model, 
the origin of the mass hierarchy exists in 
the ``compositeness''. 
This model naturally explains why 
the large flavor mixing 
is realized in the lepton sector while
the small mixing is realized in the quark sector.
This miracle comes from the field 
contents of $SU(5)$ in 
Eq.(\ref{miracle}). 
When ${\bf 10}_i$ and ${\bf 1}_i$ of 
$SU(5)$ produce the mass 
hierarchy, 
the large (small) mixing is realized 
in the lepton (quark) sector.

\par
Now we discuss whether 
this model can solve the solar neutrino problem, or not. 
We can see that 
three mass eigen values of light neutrinos are all of 
order $\mu$ from Eq.(\ref{62}). 
Then, two mass squared differences are naively
of order 
$\delta m^2 \simeq \mu^2 \simeq 10^{-2} {\rm eV}^2$. 
Thus, 
we must introduce small parameters in the coefficients 
in order to 
obtain the mass squared difference 
of $O(10^{-5})$ eV$^2$ for the MSW solution, or 
$O(10^{-10})$ eV$^2$ for the vacuum oscillation 
solution. 
Here we assume that 
the neutrino mass matrix Eq.(\ref{62}) 
has rank one
\footnote{ 
We can also consider the situation 
that three mass eigen values of $O(\mu)$ are closely 
degenerated and have small mass squared 
differences of $O(10^{-5})$ eV$^2$ 
or $O(10^{-10})$ eV$^2$.
This situation also needs small 
parameters in the coefficients of 
neutrino mass matrix. 
}, 
which is so-called ``democratic type'' mass matrix. 
In this case, 
three mass eigen values become of $O(0)$, $O(0)$, and 
$O(\mu)$. 
In order to 
solve the solar neutrino problem, 
we must introduce small parameters in the coefficients 
of Eq.(\ref{62}) as the mass perturbation\cite{10.5}.

\par
Here we should comment on $R$-parity, 
which distinguishes $\overline{Q_i}$ from 
$\overline{H}$, 
and $N_i$ from $S$. 
It is difficult to introduce $R$-parity 
at the preon level\footnote{
One of the simplest example is the introduction of 
$Z_4$ symmetry. 
We assign $Z_4$ charge as 
$Q(i)$, $Q'(-i)$, $A(+)$, $N_3(-)$, 
$\overline{\chi_j}(+)$, $\overline{Q_i}(-)$, 
$H(+)$, $\overline{H}(+)$, and $S(+)$. 
However, $N_1$ and $N_2$ have wrong signs in 
this case. 
}. 
$N_i$s can be easily distinguished from $S$ 
by $U(1)_w$ gauge symmetry. 
On the other hand, for $\overline{\bf 5}$ fields, 
we simply assume that operators 
$[ \overline{Q_i} H S]$, 
$[ \overline{H} H S {\bf 1}_i ]$, and 
$[{\bf 10}_{i_1}\: \overline{Q_{i_2}}\: \overline{Q_{i_3}}]$ 
are forbidden. 
The absence of operator which is 
consistent with all symmetries 
sometimes happens in string derived models. 
%As we have assumed before, 
%$\chi_j$ and $\overline{\chi_j}$ are 
%heavy and integrated out around the GUT scale. 
Then the conventional $R$-parity symmetry 
as well as $U(1)_{B-L}$ global symmetry 
appear in the effective theory bellow the 
confinement scale.

\par
We can also derive the same structures of mass matrices as 
Eqs.(\ref{up2})$\sim$(\ref{neuD2}) by another mechanism. 
One example is represented in the Appendix A, 
where the origin of mass hierarchy is produced 
not by the ``compositeness'' but by the discrete symmetry.

\par
We would like to show another possibility 
before closing this section. 
We can obtain 
the large (small) mixing in the 
lepton (quark) sector, 
even in the case that 
only ${\bf 10}_i$ produces the mass hierarchy. 
In this case, contrary to our model, 
there is no mass hierarchy 
in the Dirac mass and Majorana mass of neutrinos, 
since ${\bf 1}_i$ does not produce 
mass hierarchy. 
The model with three sets of vector-like extra generations
of {\bf 10} and $\overline{\bf 10}$ 
which is 
built by Babu and Barr is just the case\cite{Barr}.
Another example is the composite model 
built by Strassler\cite{Strassler}, 
where the possibility of the 
large (small) mixing in the lepton (quark) 
sector have been mentioned 
by Strassler and Yanagida\cite{Strassler2}.

\par
We can easily build the composite model 
which induces the same results of Babu and Barr 
based on Ref.\cite{Strassler}\footnote{
In the original model in Ref.\cite{Strassler}, 
top Yukawa of $O(1)$ is generated dynamically. 
Here we consider the situation that 
top Yukawa is also 
induced from perturbative interaction 
by introducing elementary Higgs fields. 
The composite ``Higgs field'' in 
Ref.\cite{Strassler} corresponds to 
$\chi_j$s in our model. 
}.
Here we show it briefly. 
We introduce the hyper-color of 
$Sp(2N)$ with $N_f = N + 2$ $(N>1)$ fundamental 
representations for each generation. 
Contrary to our composite model, 
there are three hyper-colors in 
addition to the standard gauge group. 
The composite state of ${\bf 10}_i$ 
induces the hierarchy parameter
$\epsilon_i \equiv \Lambda_i/M_p$ $(i=1,2,3)$, 
and the mass hierarchy of quark and lepton 
is given by 
\begin{eqnarray}
& & m_d : m_s : m_b \sim m_e : m_{\mu} : m_{\tau} \sim 
    \epsilon_1 :  \epsilon_2 :  \epsilon_3 \;,  \nonumber \\
& & m_u : m_c : m_t \sim \epsilon_1^2 :  
    \epsilon_2^2 :  \epsilon_3^2 \; .
\end{eqnarray}
Since the proton stability demands 
$10^{-2}< \epsilon_i < 1$,
we consider $\epsilon_3 \sim 1$ and 
$\epsilon_i/\epsilon_{i+1} \simeq 10^{-1}$.
It is the model of large $\tan \beta$. 
This model also induces the large mixing 
in the lepton sector as 
$V^{\rm lepton}_{}{}_{ij} \simeq 1$ 
and the small mixing
in the quark sector as 
$V^{\rm quark}_{KM}{}_{ij} \simeq \epsilon_i/\epsilon_j$.

%
%
%%%%%%%%%%%%%%%%%%%%%%%%%%%%%%%%%%%%%%%%%%%%%%%%%%%%%%%%%%%%%%%%%%%%%%%%%
%%%%%%%%%%%%%%%%%%%%%%%%%%%%%%%%%%%%%%%%%%%%%%%%%%%%%%%%%%%%%%%%%%%%%%%%%
%%%%%%%%%%%%%%%%%%%%%%%%%%%%%  Section 3  %%%%%%%%%%%%%%%%%%%%%%%%%%%%%%%
%%%%%%%%%%%%%%%%%%%%%%%%%%%%%%%%%%%%%%%%%%%%%%%%%%%%%%%%%%%%%%%%%%%%%%%%%
%%%%%%%%%%%%%%%%%%%%%%%%%%%%%%%%%%%%%%%%%%%%%%%%%%%%%%%%%%%%%%%%%%%%%%%%%
%
\section{Improving the model}

The previous model 
naturally induces 
the large (small) mixing 
in the lepton (quark) sector. 
We can naturally solve the atmospheric 
neutrino problem. 
However, 
we must introduce small parameters 
in the coefficients of neutrino mass matrix 
in order to 
solve the solar neutrino problem. 
In this section we try to improve the model 
in order to obtain the natural solution of 
the solar neutrino problem.

\par
{}For this purpose, 
we introduce the discrete symmetry $Z_2'$ 
and one more gauge singlet field $\Phi$.  
Under this discrete symmetry 
only ${\overline{\bf 5}_1}$ and 
$\Phi$ have odd charges, while 
other fields have even charges. 
We assume that $\Phi$ takes VEV as 
$\langle \Phi \rangle/M_p \ll 1$\footnote{ 
$Z_2$ symmetry is broken at the scale of 
$\langle \Phi \rangle$. 
However, mass matrices of quark and lepton 
do not change 
since we consider the case of 
$g \langle \Phi \rangle/M_p \ll 1$.
}.  
Then the mass matrices of $m^{d}_{ij}$, 
$m^{e}_{ij}$, and $m^{\nu}_{ij}$ in 
Eqs.(\ref{down2}), (\ref{lepton2}), and (\ref{neuD2}) 
are modified as 
\begin{equation}
\label{down2p}
m^{d}_{ij} \simeq 
 \left(
\begin{array}{ccc}
 \phi \: \epsilon^2   &  \epsilon^2   &  \epsilon^2    \\
 \phi \: \epsilon     &  \epsilon     &  \epsilon      \\
 \phi              &  1            &  1             \\
\end{array}
 \right) \: \overline{v}\; ,
\end{equation}
\begin{equation}
\label{lepton2p}
m^{e}_{ij} \simeq 
 \left(
\begin{array}{ccc}
 \phi \: \epsilon^2   &  \phi \: \epsilon     &  \phi     \\
 \epsilon^2   &  \epsilon     &  1     \\
 \epsilon^2   &  \epsilon     &  1     \\
\end{array}
 \right)  \: \overline{v}\; ,
\end{equation}
\begin{equation}
\label{neuD2p}
m^{\nu}_{ij} \simeq 
 \left(
\begin{array}{ccc}
 \phi \: \epsilon^2   & \phi \: \epsilon   & \phi    \\
  \epsilon^2   & \epsilon   &  1   \\
  \epsilon^2   & \epsilon   &  1   \\
\end{array}
 \right) \: g^2 \: v  \;,
\end{equation}
respectively. 
Where we define 
$\phi \equiv g \langle \Phi \rangle/M_p \ll 1$. 
$m^{u}_{ij}$ and $M_{\nu}$ have the same form as 
Eq.(\ref{up2}) and Eq.(\ref{majo22}), respectively. 
The mass matrix of three light 
neutrinos $m_{\nu_l}$ becomes 
\begin{eqnarray}
\label{62p}
& &  m_{\nu_l} 
  = - m_{\nu} M_{\nu}^{-1} m_{\nu}{}^T , \nonumber \\
& & \simeq  
 - \left( 
    \begin{array}{ccc}
 \phi \: \epsilon^2   & \phi \: \epsilon   & \phi    \\
  \epsilon^2   & \epsilon   &  1   \\
  \epsilon^2   & \epsilon   &  1   \\
   \end{array}
  \right) 
  \left(
\begin{array}{ccc}
1/ \epsilon^4          &  1/ \epsilon^3   &  1/ \epsilon^2   \\
1/ \epsilon^3          &  1/ \epsilon^2   &  1/ \epsilon     \\
1/ \epsilon^2          &  1/ \epsilon     &       1          \\
\end{array}
   \right) 
   \left( 
    \begin{array}{ccc}
 \phi \: \epsilon^2   & \epsilon^2   & \epsilon^2    \\
 \phi \: \epsilon     & \epsilon     & \epsilon      \\
 \phi               & 1            &  1            \\
    \end{array}
  \right) {g^4 v^2 \over M_p}, \nonumber \\
& & \simeq  
 - \left( 
 \begin{array}{ccc}
  \phi^2 & \phi & \phi \\
  \phi   & 1    & 1    \\
  \phi   & 1    & 1    \\
 \end{array}
 \right) {g^4 v^2 \over M_p} .  
\end{eqnarray}

\par
Let us estimate the flavor mixing in the quark and 
the lepton sector. 
{}For the quark sector, the flavor mixing 
matrix Eq.(\ref{KMq}) does not change. 
$V_{cb}$ and $V_{ub}/V_{cb}$ are the same as 
in the Eq.(\ref{VKMq}), which are roughly consistent 
with experiments. 
On the other hand, 
$V_{us} \simeq \epsilon$ becomes larger than 
the value of ration ${m_d/m_s}$, 
because Eq.(\ref{ratio2}) is modified as  
\begin{eqnarray}
\label{ratio2p}
& & m_d : m_s : m_b \sim  m_e : m_{\mu} : m_{\tau} \sim 
    \phi \ \epsilon^2 : \epsilon : 1  \; , \nonumber \\ 
& & m_u : m_c : m_t \sim 
    \epsilon^4 : \epsilon^2 : 1 \;.
\end{eqnarray}
If $\phi \simeq O(10^{-1})$, 
the value of $V_{us}$ and 
the mass hierarchy of the first and 
the second generations of down-sector and 
charged lepton become realistic.

\par
As for the lepton sector, 
unitary matrices $U_L^e$ and $U^{\nu}$ in Eq.(\ref{UU}) 
are modified as 
\begin{equation}
\label{UUp}
 U_L^e \sim U^{\nu} \simeq  \left( 
 \begin{array}{ccc}
  1      & \phi           & \phi             \\
  \phi   & \cos \theta    & - \sin \theta    \\
  \phi   & \sin \theta    & \cos \theta      \\
 \end{array}
\right) ,
\end{equation}
where $\theta$ is a mixing angle of $O(1)$.
It suggests that 
$V^{\rm lepton}_{}\equiv U_L^e{}^{\dagger}U^{\nu}$ 
has the same form as Eq.(\ref{UUp}).
This form seems to be suitable for 
the small mixing solar neutrino MSW solution of 
$\nu_{e}$ and $\nu_{\mu}$, 
and the large mixing atmospheric neutrino solution 
of $\nu_{\mu}$ and $\nu_{\tau}$. 
However, it is not true. 
Equation (\ref{UUp}) is derived because 
we estimate masses of three light 
neutrinos as 
$O(\phi^2 \mu)$, $O(\mu)$, and $O(\mu)$ in Eq.(\ref{62p}). 
Since two mass squared differences are both of 
$O(\mu^2)$ in this case, we can not solve 
the solar neutrino problem without 
introducing small parameters 
in the coefficients of neutrino mass matrix 
Eq.(\ref{62p}).

\par
Here we assume that the determinant of 
$2 \times 2$ sub-matrix of the second and 
the third generations is zero at order $\mu^2$ 
in Eq.(\ref{62p})\footnote{
This situation is also the 
fine-tuning at this stage. 
We expect this situation might be 
realized by the 
string derived model. 
}. 
In this case, 
three mass eigen values 
are of $O(\phi \mu)$, $O(\phi \mu)$, 
and $O(\mu)$. 
The unitary matrix $U^{\nu}$ 
is modified as 
\begin{equation}
\label{UUpp}
 U^{\nu} \sim \left( 
 \begin{array}{ccc}
  1/\sqrt{2}   & 1/2           &  1/2        \\
  -1/\sqrt{2}  & 1/2           &  1/2        \\
  \phi         & -1/\sqrt{2}   &  1/\sqrt{2} \\
 \end{array}
\right) .
\end{equation}
This form is justified for the sufficient small $\phi$. 
Then the lepton flavor mixing 
$V^{\rm lepton}_{}\equiv U_L^e{}^{\dagger}U^{\nu}$ 
can induce the large mixings of $(\nu_{e}- \nu_{\mu})$ and 
$(\nu_{\mu}- \nu_{\tau})$\footnote{
To be accurate, 
$\theta \simeq O(1)$ of $U_L^e$ in Eq.(\ref{UUp}) 
should be sufficient small 
in order to realize ``bi-maximal mixing''. 
This situation can be easily realized 
by suitable coefficients of 
charged lepton mass matrix. 
}.
This case is so-called ``bi-maximal mixing''. 
If we input $\phi \simeq 10^{-1.5}$, 
we obtain 
$\delta m^2_{12} \simeq 10^{-5}$ eV$^2$, 
which is nothing but 
the large angle MSW solution.  
Besides, this case suggests 
the realistic 
mass hierarchy of quark and lepton, 
and realistic value of $V_{us}$ as we have 
seen before. 
Thus, we can solve not only atmospheric 
neutrino problem but also 
the solar neutrino problem 
by the large angle MSW solution 
in this model. 
On the other hand, if we input 
$\phi \simeq 10^{-4}$, 
we obtain 
$\delta m^2_{12} \simeq 10^{-10}$ eV$^2$, which 
is suitable for 
the vacuum oscillation solutions\footnote{
The ``bi-maximal mixing'', which 
explains the atmospheric neutrino oscillation 
and the vacuum oscillation for the solar 
neutrino have been studied 
in Refs.\cite{12}\cite{14}\cite{YT}.
}. 
Unfortunately, masses of electron and down quark 
in Eq.(\ref{ratio2p}) become too small 
when $\phi \simeq 10^{-4}$. 
This case might also be a realistic solution 
by extending the model.

%
%%%%%%%%%%%%%%%%%%%%%%%%%%%%%%%%%%%%%%%%%%%%%%%%%%%%%%%%%%%%%%%%%%%
%%%%%%%%%%%%%%%%%%%%%%%%%%%%%%%%%%%%%%%%%%%%%%%%%%%%%%%%%%%%%%%%%%%
%%%%%%%%%%%%%%%%%%%%%%%  Summary    %%%%%%%%%%%%%%%%%%%%%%%%%%%%%%%
%%%%%%%%%%%%%%%%%%%%%%%%%%%%%%%%%%%%%%%%%%%%%%%%%%%%%%%%%%%%%%%%%%%
%%%%%%%%%%%%%%%%%%%%%%%%%%%%%%%%%%%%%%%%%%%%%%%%%%%%%%%%%%%%%%%%%%%
%
\section{Summary and Discussion}

In this paper we suggest a composite model
which can naturally induce the large flavor mixing
in the lepton sector and the 
small mixing in the quark sector.
This model has only one hyper-color in addition
to the standard gauge group, 
which makes composite states of preons. 
In this model, {\bf 10} and {\bf 1} representations 
in $SU(5)$ are composite states, and they produce 
the mass hierarchy. 
This can explain why the large mixing is realized 
in the lepton sector, while the small mixing 
is realized in the 
quark sector. 
This model can naturally solve 
the atmospheric 
neutrino problem. 
In the improved model, 
we can derive the mass scale which is 
suitable for the solution of 
the solar neutrino problem.  
We can solve 
not only atmospheric 
neutrino problem but also 
the solar neutrino problem 
by the large angle MSW solution 
in this model. 
The vacuum oscillation solution 
might be possible by extending the model.

\vskip 1 cm
\noindent
{\large{\bf Acknowledgements}}\par
I would like to thank Prof. T. Yanagida 
{}for many useful discussions 
and many fruitful comments. 
I would like to thank Profs. A. I. Sanda, M. Tanimoto,
T. Matsuoka, M. Suzuki, H. Murayama, 
C. S. Lim, and K. Inoue for 
useful discussions and comments. 
I would like to thank Profs. V. A. Miransky and Y. Sugiyama for 
careful reading of manuscript.

%
%
%
%
%%%%%%%%%%%%%%%%%%%%%%%%%%%%%%%%%%%%%%%%%%%%%%%%%%%%%%%%%%%%%%%%%%%%%%%%%%
%%%%%%%%%%%%%%%%%%%%%%%%%%%%%%%%%%%%%%%%%%%%%%%%%%%%%%%%%%%%%%%%%%%%%%%%%%
%%%%%%%%%%%%%%%%% Appendix %%%%%%%%%%%%%%%%%%%%%%%%%%%%%%%%%%%%%%%%%%%%%%%
%%%%%%%%%%%%%%%%%%%%%%%%%%%%%%%%%%%%%%%%%%%%%%%%%%%%%%%%%%%%%%%%%%%%%%%%%%
%%%%%%%%%%%%%%%%%%%%%%%%%%%%%%%%%%%%%%%%%%%%%%%%%%%%%%%%%%%%%%%%%%%%%%%%%%
\appendix
\section{A model with discrete symmetry}

Here we suggest the model where the discrete symmetry 
plays a crucial role of producing mass hierarchy, which induces 
the large (small) mixing of the lepton (quark) sector. 
We introduce a gauge group $Sp(8)$ 
and its ten fundamental representations $P$s.
We also introduce the discrete symmetry $Z_5 \times Z_2''$.
The representations of fields 
are the followings.
\begin{equation}
 \begin{array}{c|ccc}
  {\rm field}   &   Sp(8)      &    SU(5)  & Z_5 \times Z_2''     \\ 
\hline
  & & & \\
  P             &   \Yfund     &      1    & ( \omega,   -)       \\
{\bf 10}_3      &     1        &  \Yasymm  & ( 1,        +)       \\
{\bf 10}_2      &     1        &  \Yasymm  & ( \omega^3, +)       \\
{\bf 10}_1      &     1        &  \Yasymm  & ( \omega,   +)       \\
{\overline{\bf 5}_3}& 1        & \overline{\Yfund} & ( \omega^3, +) \\
{\overline{\bf 5}_2}& 1        & \overline{\Yfund} & ( \omega^3, +) \\
{\overline{\bf 5}_1}& 1        & \overline{\Yfund} & ( \omega^3, +) \\
{\bf 1}_3       &     1        &      1    & ( \omega^2, +)       \\
{\bf 1}_2       &     1        &      1    & ( \omega  , +)       \\
{\bf 1}_1       &     1        &      1    & ( \omega^4, +)       \\
  H             &     1        &   \Yfund  & ( 1,        +)       \\
 \overline{H}   &     1        & \overline{\Yfund} & (1, +)       \\ 
 \end{array}
\end{equation}  
The $Sp(8)$ hyper-color makes the composite 
state $M \equiv PP$.
The non-perturbative effects of $Sp(8)$ 
gauge symmetry induce the superpotential\cite{Sp}
\begin{equation}
 W_{dyn} = X({\rm Pf}M - \Lambda^{10}) ,
\end{equation}
which makes the vacuum 
$\langle M \rangle \simeq \Lambda^2$. 
$\Lambda$ is the strong coupling scale of $Sp(8)$. 
$Z_5 \times Z_2''$ symmetry reduces to $Z_5$ symmetry 
bellow the confinement scale $\Lambda$.

\par
Let us show the mass matrices of quark and lepton.
Mass terms which are not singlet under 
the discrete symmetry $Z_5$ 
are produced from the irrelevant operators
suppressed by the Planck scale $M_p$. 
The mass hierarchy is expressed by the 
small dimension-less parameter 
$\eta \equiv \langle M \rangle /M_p^2 \ll 1$. 
The mass matrix of the up-sector $m^{u}_{ij}$, 
down-sector $m^{d}_{ij}$, 
charged lepton $m^{e}_{ij}$, and 
Dirac neutrino $m^{\nu}_{ij}$ are given by 
\begin{equation}
\label{up1}
m^{u}_{ij} \simeq 
 \left(
\begin{array}{ccc}
     \eta^4     &   \eta^3    &  \eta^2    \\
     \eta^3     &   \eta^2    &  \eta      \\
     \eta^2     &   \eta      &   1        \\
\end{array}
 \right) \: v  ,
\end{equation}
\begin{equation}
\label{down1}
m^{d}_{ij} \simeq 
 \left(
\begin{array}{ccc}
 \eta^3   & \eta^3  & \eta^3    \\
 \eta^2   & \eta^2  & \eta^2    \\
 \eta     & \eta    & \eta      \\
\end{array}
 \right)  \: \overline{v} , 
\end{equation}
\begin{equation}
\label{lepton1}
m^{e}_{ij} \simeq 
 \left(
\begin{array}{ccc}
 \eta^3   & \eta^2  & \eta      \\
 \eta^3   & \eta^2  & \eta      \\
 \eta^3   & \eta^2  & \eta      \\
\end{array}
 \right)  \: \overline{v}  ,
\end{equation}
\begin{equation}
\label{neuD1}
m^{\nu}_{ij} \simeq 
 \left(
\begin{array}{ccc}
 \eta^4   & \eta^3  & 1      \\
 \eta^4   & \eta^3  & 1      \\
 \eta^4   & \eta^3  & 1      \\
\end{array}
 \right)  \: v  ,
\end{equation}
respectively. 
The Majorana mass of right-handed neutrinos $M_{\nu}$
is given by
\begin{equation}
\label{majo1}
M_{\nu} \simeq 
\left(
\begin{array}{ccc}
 \eta     &  1      & \eta^2      \\
  1       & \eta^4  & \eta        \\
 \eta^2   & \eta    & \eta^3      \\
\end{array}
\right) \; M_p .
\end{equation}
The mass matrix of three light 
neutrinos $m_{\nu_l}$ is given by 
\begin{eqnarray}
\label{61}
& &  m_{\nu_l} 
  = - m_{\nu} M_{\nu}^{-1} m_{\nu}{}^T , \nonumber \\
& & \simeq  
 - \left( 
    \begin{array}{c}
     1  \\
     1  \\
     1  \\
   \end{array}
  \right) 
\left(
\begin{array}{ccc}
  \eta^4  &  \eta^3  &  1    \\
\end{array}
\right) 
\left(
\begin{array}{ccc}
1/ \eta      &  1          &  1/ \eta^2   \\
1            &  \eta       &  1/ \eta     \\
1/ \eta^2    &  1/ \eta    &  1/ \eta^3   \\
\end{array}
\right) 
\left( 
    \begin{array}{c}
        \eta^4   \\
        \eta^3   \\
          1      \\
   \end{array}
  \right)
\left(
\begin{array}{ccc}
  1 & 1 & 1 \\
\end{array}
\right) {v^2 \over M_p} , \nonumber \\ 
& & \simeq  
 - \left( 
 \begin{array}{ccc}
  1 & 1 & 1 \\
  1 & 1 & 1 \\
  1 & 1 & 1 \\
 \end{array}
 \right) {v^2 \over M_p \eta^3} \; .  
\end{eqnarray}

\par
The following mass hierarchy is derived 
in this model. 
\begin{eqnarray}
\label{ratio1}
& & m_d : m_s : m_b \sim  m_e : m_{\mu} : m_{\tau} \sim 
    \eta^3 : \eta^2 : \eta  \nonumber \\ 
& & m_u : m_c : m_t \sim \eta^4 : \eta^2 : 1 
\end{eqnarray}
This model suggests small $\tan \beta$. 
This model 
naturally induces the large mixing 
in the lepton sector 
and the small mixing 
in the quark sector. 
It is because 
${\bf 10}_i$ and ${\bf 1}_i$ 
produce the mass hierarchy\footnote{
Three ${\overline{\bf 5}_i}$s have 
the same discrete charge while 
${\bf 10}_i$s and ${\bf 1}_i$s 
have different charges corresponding to the 
generation.
} 
as the composite model presented in this paper.
Here the mass hierarchy is produced not 
by the ``compositeness'' but 
by the discrete symmetry.

%
%%%%%%%%%%%%%%%%%%%%%%%%%%%%%%%%%%%%%%%%%%%%%%%%%%%%%%%%%%%%%%%%
%%%%%%%%%%%%  References  %%%%%%%%%%%%%%%%%%%%%%%%%%%%%%%%%%%%%%%
%%%%%%%%%%%%%%%%%%%%%%%%%%%%%%%%%%%%%%%%%%%%%%%%%%%%%%%%%%%%%%%%%

\end{document}